\documentclass[layout=twocolumn]{achemso}
\setkeys{acs}{keywords=true} 
\usepackage{graphicx}
\usepackage{dcolumn}
\usepackage{bm}
\usepackage{hyperref}
\usepackage{xcolor}
\usepackage{comment}
\usepackage{amsmath}

\title{Manipulating the Stacking in Two-Dimensional Hexagonal Boron Nitride Bilayers: Implications for Defect-Based Single Photon Emitters
}

\author{Fernanda Pinilla}
\affiliation{Departamento de F\'isica, Facultad de Ciencias, Universidad de Chile, 7800024 \~Nu\~noa, Santiago, Chile}

\author{Wilver A. Muriel}
\affiliation{Departamento de F\'isica, Facultad de Ciencias, Universidad de Chile, 7800024 \~Nu\~noa, Santiago, Chile}

\author{Javiera Cabezas-Escares}
\affiliation{Departamento de F\'isica, Facultad de Ciencias, Universidad de Chile, 7800024 \~Nu\~noa, Santiago, Chile}

\author{Ignacio Chac\'on}
\affiliation{Departamento de F\'isica, Facultad de Ciencias, Universidad de Chile, 7800024 \~Nu\~noa, Santiago, Chile}

\author{Carlos Cardenas}
\email{cardena@uchile.cl}
\affiliation{Departamento de F\'isica, Facultad de Ciencias, Universidad de Chile, 7800024 \~Nu\~noa, Santiago, Chile}
\alsoaffiliation{Center for the Development of Nanoscience and Nanotechnology (CEDENNA), 9170124 Santiago, Chile}

\author{Francisco Munoz}
\email{fvmunoz@gmail.com}
\affiliation{Departamento de F\'isica, Facultad de Ciencias, Universidad de Chile, 7800024 \~Nu\~noa, Santiago, Chile}
\alsoaffiliation{Center for the Development of Nanoscience and Nanotechnology (CEDENNA), 9170124 Santiago, Chile}

\date{\today}

\begin{document}


\begin{abstract}

{\color{black}Defect-based single photon emitters are a very interesting option for quantum technologies. Unlike bulk-based emitters, two-dimensional materials provide advantages, such as the possibility of forming van der Waals heterostructures. In them, the adjacent layers can be used to control or modify some properties, often in a controlled way. In this work, we show that the emission energy of single emitters is affected by the potential exerted by an adjacent insulating layer. We applied this insight to the hexagonal boron nitride bilayer, and depending on the actual defect and stacking order, the emission energy can be changed in the range $\sim 10-400$ meV.}

\end{abstract}

\maketitle

\section{\label{sec:intro}Introduction}

{\color{black}Single-photon emitters (SPEs) with exceptional brightness, high fidelity, and efficient coupling to external fields are crucial building blocks for next-generation quantum technologies like quantum cryptography and ultraprecise sensors. Although well-established emitters such as the diamond NV center offer valuable functionalities, their intrinsic limitations, such as limited photon emission rates,\cite{Aharonovich_2011} challenging integration with nanophotonic structures,\cite{Kim:20} and limited control over emission properties,\cite{Bathen2021} hinder their broader applicability. 

Recent advancements have thrust two-dimensional (2D) materials into the spotlight as promising hosts for SPEs.\cite{Azzam2021} In these atomically thin layers, luminescent point defects, such as those found in transition metal dichalcogenides (TMD) and hexagonal boron nitride (hBN), reside at the same shallow depth, inherently overcoming bulkiness limitations and enabling direct manipulation of their optical properties with external fields. This opens avenues for unprecedented control over light emission and efficient integration with nanophotonic circuits, paving the way for new miniaturized and highly scalable quantum devices. However, 2D SPEs have practical problems, such as blinking (intermittence) or bleaching (fading) of the emission.\cite{Martinez2016} Among the most promising ways to control the emission of SPEs hosted by 2D materials, is by stacking layers in van der Waals (vdW) heterostructures.\cite{geim2013van} In these systems, the individual layers mutually influence each other, often bringing new functionality to the entire structure.} Recently, White \textit{et al.} designed a heterostructure based on hBN multilayers, hosting SPEs and graphene.\cite{white21} By applying a transversal voltage, they successfully toggled specific SPEs on/off, demonstrating precise control over the filling of their electronic levels.

{\color{black}The properties induced by a vdW heterostructure can often be understood in terms of simple models or mechanisms. For instance, regardless of the actual luminescent point defect (\textit{e.g.} C-based, vacancy, etc.), they are expected to have a non-zero dipole moment at least in one of the levels involved in the transition. There should then be an interaction between the dipole moment of a SPE and the dipolar texture of an adjacent layer, if present. This interaction is expected to change the emission energy, and exploring these changes is the focus of this article. To keep the picture simple, we will consider both layers as insulators with their band gaps properly aligned, at least in regions close to the defect levels involved in the transition. Although this model is broadly applicable, our focus will be on the hBN bilayer. This choice offers several advantages for theoretical description, including a diverse range of SPEs and several different external potentials due to different stacking arrangements.} While some stacking orders can be induced during synthesis,\cite{Gilbert2019} other arrangements can be experimentally induced by relative sliding or twisting among the layers,\cite{Yasuda21, Vizner2021, Jiang2022}  and even a ferroelectric state can be achieved.

We start by mentioning our calculation methods. Then, we present a simple {\color{black} theoretical picture to convince the reader that such a modification in the emission energy is expected.} {\color{black}In this section, our discussion will focus on providing a general understanding of the effect, but without delving into specific materials or their corresponding emitters, the limitations of this model will be discussed}. Next, we show our results -calculated with density functional theory- for the bare bilayer and for the bilayer hosting SPEs. The validity of our simple model is explored. Finally, we present our conclusions.

\section{\label{sec:methods}Computational Methods}

Calculations were performed using density functional theory (DFT) with the VASP package\cite{vasp1,vasp2,vasp3,vasp4}. The geometries were relaxed using the Perdew–Burke-Ernzerhof\cite{pbe} (PBE) exchange-correlation functional, while the defect states and optical transitions were calculated using the Heyd-Scuseria-Ernzerhof (HSE) hybrid functional\cite{hse03,hse06}. This scheme has previously been used.\cite{Auburger21,jara2021} The Tkatchenko-Scheffler method \cite{Tkatchenko09} was used to account for the dispersion energy and obtain the distance between layers for each hBN arrangement. This produced an interlayer distance of 3.34~\AA, in excellent agreement with the experimental value of 3.33~\AA. Excited states were studied using the $\Delta$SCF method,\cite{Jin21} it consist on simulating the excitation by manually setting the occupation of the levels involved. {\color{black}Within this approximation, the energies of excited singlet states are not straightforward, as DFT calculations provide mixed-spin single states (see the supplementary information of Ref.\cite{Mack2019}). In this article, we used these mixed-spin excited singlets without correcting them, as their energetics are similar to those of other studies applying such corrections.\cite{Auburger21} The accuracy of HSE for defect calculations has been studied in the bulk\cite{Deak2010}. However, the reduced screening in 2D materials does not guarantee that accuracy is transferable. In this work our main interest is not on the absolute values of the defect levels, but on their changes in energy induced by changing the stacking of a van der Waals system; such changes should be accurately described, as explained our theory framework. }

Regarding the other calculation parameters and settings, projector-augmented wave (PAW) type pseudopotentials \cite{paw} were used and the kinetic energy cutoff was set to 400 eV. {\color{black} A vacuum layer of $\sim 15$\AA was added to simulate the two-dimensional geometry.} We tested a higher cutoff (650 eV), which did not significantly alter the results. A single k-point ($\Gamma$) was used in the supercell calculations. For bulk calculations, a grid of $15\times 15$ k-points was used. Results analysis was performed using PyProcar \cite{pyprocar,LANG2024109063}, and visualization was performed with VESTA \cite{vesta}.

All calculations shown here use a $7\times 7$ supercell{\color{black}, \textit{i.e.} 98 atoms}.  In some cases, a $8\times 8$ supercell was tested, giving similar results. For the calculation of defects, the distance between layers was fixed at $3.34$~\AA.  The system was allowed to fully relax in the in-plane coordinates, and the atomic rearrangements resulting from excitation involved only in-plane displacements for the defects studied.

\section{Modulation of the gap with the stacking order}
\label{sec:model}

\label{sec:generalModel}

Conceptual DFT\cite{geerlings2003,fuentealba2013} provides a neat framework for  {\color{black}understand the changes in energy of defect levels within the fundamental band gap of a wide band gap insulator} using the landscape of the \textit{external} potential, $v(r)$, from {\color{black}adjacent layers in a vdW heterostructure}.  The Janak theorem\cite{Janak} relates the $i$-th Khon-Sham eigenvalue $\varepsilon_i$ to the derivative of the total energy $E$ with the occupation number of that orbital $n_i$,
\begin{equation}
    \varepsilon_i = \left(\frac{\partial E}{\partial n_i}\right)_{v(r)}.
\end{equation}
{\color{black}Here, the states of interest are the defect levels, which we will assume to be within the fundamental band gap of the heterostructure, that is, they are the highest occupied and lowest unoccupied molecular orbitals, or HOMO (H) and LUMO (L), respectively.} The linear variation (\textit{small} change) of the HOMO-LUMO gap, $\Delta\varepsilon_{HL-gap}$, due to the external potential, would be,
\begin{align}
\Delta \varepsilon_{HL-gap} &= \int \frac{\delta}{\delta v(r)} \left( \frac{\partial E}{\partial n_{L}} - \frac{\partial E}{\partial n_{H}} \right) \delta v(r) dr \\
&= \int \left( \frac{\partial \rho(r)}{\partial n_{L}} - \frac{\partial \rho(r)}{\partial n_{H}} \right) \delta v(r) dr.
\end{align}
where $\rho(r)$ is the electron density, and we have used the fact that $\left(\frac{\delta E}{\delta v(r)}\right)=\rho(r)$.  The derivatives of the density with respect to HOMO/LUMO are the Fukui functions\cite{ceron2020}. However,   if one neglects the small relaxation of inner states as the occupation of HOMO and LUMO is varied, these derivatives are simply the ``densities'' of these orbitals, $|\phi_H(r)|^2$ and $|\phi_L(r)|^2$, such that 
\begin{equation}
\label{eq:gap1}
    \Delta \varepsilon_{HL-gap} \approx \int \left(  |\phi_L(r)|^2  - |\phi_H(r)|^2  \right )\delta v(r) dr.   
\end{equation}
{\color{black}Eq.~\ref{eq:gap1} is the main result of this section: the energy gap between the emitter levels will be affected by the vdW heterostructure if the average potential between both levels is different. Regardless of the details of the actual defect, the densities of the ground and excited states are expected to be different. In the remainder of this section, we will try to gain insight into $\Delta\varepsilon_{HL-gap}$ for a particularly simple type of defects.}

In {\color{black} some SPEs}, the HOMO and LUMO are well localized around the defect atoms/vacancies {\color{black}and their closest neighbors}.  Hence, {\color{black}as a first approximation}, let us assume that {\color{black}\textit{(i)} each orbital is well localized within a domain $\Omega_i$ of volume $V(\Omega_i)$, and \textit{(ii)} the orbital is constant inside $\Omega_i$. Then}  $|\phi_i(r)|^2\approx 1/V(\Omega_i)\  \textit{if } r \in\Omega_i$. With this, Eq.~\ref{eq:gap1} simplifies to 
\begin{align}
\Delta \varepsilon_{HL-gap} &\approx \frac{1}{V(\Omega_L)}\int_{\Omega_L} \delta v(r) dr \\
&\approx \langle \delta v(r)\rangle_{\Omega_L}-\langle \delta v(r)\rangle_{\Omega_H} \\
\label{eq:gap2}
&\approx \langle \delta\Phi(r)\rangle_{\Omega_H}-\langle \delta\Phi(r)\rangle_{\Omega_L}.
\end{align}

Here $\Phi(r)$ is the electrostatic potential produced by the pristine layer, $\langle\rangle$ denotes the spatial average{\color{black}, and labels $H,L$ in the volume $\Omega_i$ are a shorthand for HOMO and LUMO, respectively. Using projections of wave functions into atomic orbitals, the spatial average could be applied when $\Omega_i$ covers neighboring atoms. When densities $|\phi|^2_{L,H}$ are not well localized, we do not expect Eq.~\ref{eq:gap2} to have any predictive capacity; however, Eq.~\ref{eq:gap1} should remain valid as long as changes on $\phi_{H,L}$ due to $v(r)$ are small.}  

{\color{black}Finally, we discuss the limitations of the previous model. Although Eq.~\ref{eq:gap1} captures changes in defect levels due to electrostatic interaction with an adjacent layer, it ignores all other effects that occur. In particular, if we compare the defect levels of a single layer with those of a bilayer or multilayer, there is a large change in electronic screening that our model does not consider at all. Instead, our model could describe the changes in energy resulting from the different stacking orders of a bilayer or a multilayer. See Fig.~\ref{fig:stackHBN}a for an example of these changes. For example, the energy levels of the C$_2$ defect, see Fig.~\ref{fig:stackHBN}b, can be described in a tight-binding picture by the hopping strength $t$ and the on-site energies of C$_N$, C$_B$. The change from monolayer to bilayer decreases $t$, due to enhanced screening. Instead, changing the stacking order of the bilayer changes both on-site energies, and that is the effect captured by Eq.~\ref{eq:gap2}. In summary, our model describes changes in the defect's gap by changes in the electrostatic potential but does not predict the band gap of a vdW heterostructure. 

Is Eq~\ref{eq:gap1} valid for a metallic heterostructure such as hBN/TDM? Does it work for an insulating vdW heterostructure such as hBN/graphene? As long as the defect levels are within the global band gap, our model will capture the most important factor that is occurring. However, if the defect levels are within the valence or conduction bands, they will hybridize, spilling over extended states. This effect has to be added to the contribution of our model, but estimating it could be tricky. Instead of differences in external potential (Eq.~\ref{eq:gap1}), absolute changes in potential are expected to affect the spillover of defect levels. If the heterostructure is formed with a metallic layer, the metal can mediate the interaction between different atoms; for some defects with distant atoms (see Fig.~\ref{fig:stackHBN}b) that interaction can be important. For the particular case of an hBN/graphene heterostructure, changes in electrostatic potential at vdW distances should be minimal regardless of the stacking. }

Some other effects not captured by our model include the relaxation of the ions upon excitation, but this should not be a large limitation, as the gap is usually a good approximation to the ZPL. {\color{black} The Franck-Condon relaxation upon (de)excitation resonates with some phonons, producing phonon replicas in the emission spectrum. As long as \textit{(i)} the vdW heterostructure does not change these relevant phonons, and \textit{(ii)} the electron-phonon coupling is harmonic, the result should be a uniform shift in the energy of the emission spectrum, by the same amount as the ZPL. The theory behind the phonon sideband of SPEs is found in Ref. \cite{Alkauskas_2014}, it has been applied successfully to defects in hBN. Depending on the actual defect, the phonon sideband can be much more intense than the ZPL emission. Before showing the results, we will elaborate on this aspect for the SPEs of interest. 

\color{black}In addition, there are defects, such as C$_B$V$_N$,\cite{CHENG2017247}   where the densities $|\phi_{H,L}|^2$ change significantly with external potential $v(r)$. Although this invalidates Eq.~\ref{eq:gap1}, the changes in the emission energy for this particular defect are the largest found in our study.
}

\section{Results}
\label{sec:results}

\subsection{The hBN bilayer}
\label{sec:bilayer}
\begin{figure}[ht]
    \centering
    \includegraphics[width=\columnwidth]{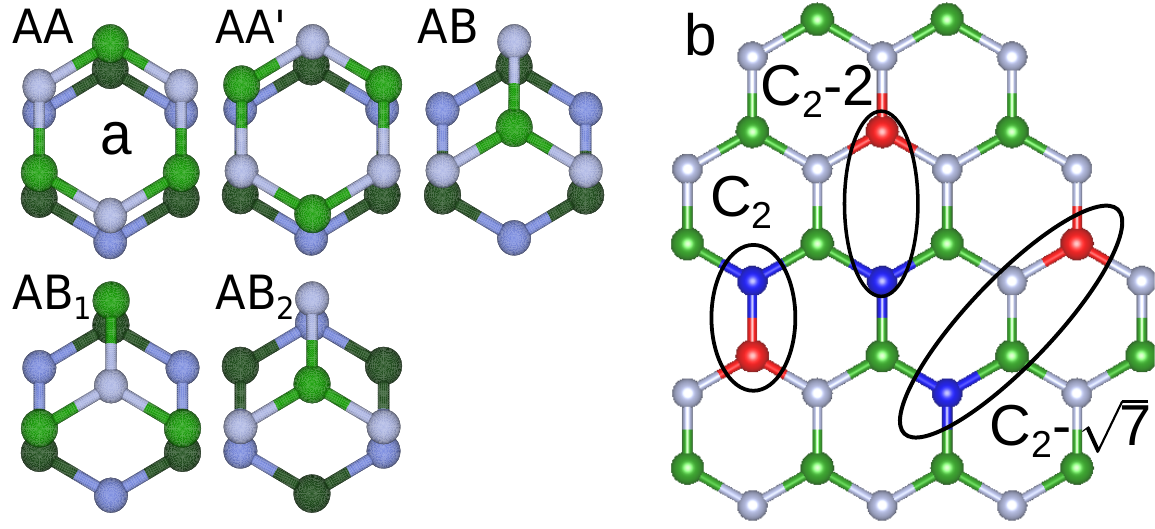}
    \caption{(a) Stacking possibilities in bilayer hBN. The B(N) atoms are denoted by green (gray) spheres{\color{black}, atoms in the lower layer have a darker color.}. In the AA (AA') stacking the B and N atoms from a layer are on top of the B(N) and N(B) atoms of the other layer, respectively. In the AB, AB$_1$, and AB$_2$ unit cell, one pair of atoms are on top of each other, but the remaining atoms from a layer lie on top of the center of a hexagonal ring from the other layer. In AB, AB$_1$, and AB$_2$ the atoms facing each other are B-N, B-B, and N-N, respectively. {\color{black}(b) Geometry of the different C$_2$-$d$ defects studied here, C$_N$ and C$_B$ are blue and red, respectively. The dimer defects are labeled by the distance between C atoms, in times the nearest neighbor distance. When C atoms are adjacent, $d=1$, we dropped the distance label.}}
    \label{fig:stackHBN}
\end{figure}

The hBN monolayer shares some resemblance with graphene; both are isoelectronic and have a similar lattice parameter. However, unlike graphene, the B and N atoms have different electronegativity, leading to a nonuniform charge pattern at the atomic level. This non-uniformity also implies the development of a large band gap in hBN. {\color{black} In our case, the dipolar texture will affect the emission energies of luminescent defects, as expected from Eq.~\ref{eq:gap2}.}

The analogy between monolayers can be extended to bilayers: surprisingly, their ground-state geometries have practically the same interlayer distance, at least for the lowest energy stacking \cite{Hod2012}. However, since both atoms of hBN are non-equivalent, there are more stacking possibilities \cite{Gilbert2019}. The AA stacking of bilayer graphene is divided into AA and AA', as shown in Fig.~\ref{fig:stackHBN}a. From simple electrostatic considerations, the AA' stacking has to be more stable than the AA stacking, since the closest interlayer neighbors have opposite charges. 

Similarly, Bernal stacking of graphene (AB) results in three possibilities, AB, AB$_1$, and AB$_2$. Among the five possible stacking orders, three of them (AA', AB, AB$_1$) have practically the same binding energy; see Table~\ref{tab:bilayer}. Two of the stacking possibilities, AA' and AB, have been synthesized.\cite{Gilbert2019}. The other arrangements could be obtained by sliding one of the layers,\cite{Yasuda21,Vizner2021, Jiang2022} or they can be found in the inversion domains (local inversion of atomic positions) \cite{Cretu2014,Li2015,Liu2012,Taha2017}.

\begin{table}[h]
    \centering
    \caption{Summary of the properties of the different stacking sequences in hBN, including the interlayer distance $c$, the Hirshfeld charges of B and N ($q_B$ and $q_N$, the charge of an electron is -1e). The energy is relative to the most abundant AA' phase and comprises a unit cell (four atoms). For comparison, the results of the literature are given in parentheses.}
    \label{tab:bilayer}
    \begin{tabular}{ccccc}
        \hline
         Stacking & $c$ (\AA) & q$_{B}$ (e) & q$_{N}$ (e) & $\Delta E$ (meV) \\\hline
         AA       & 3.53      & 0.21        & -0.21       & 25 (21\cite{Gilbert2019}, 40\cite{Constan2013}) \\\hline
         AA'      & 3.34      & 0.21        & -0.21       & 0 (0\cite{Gilbert2019}, 0\cite{Constan2013})\\\hline
         AB       & 3.36      & 0.21        & -0.21       & -3 (-2\cite{Gilbert2019}, 0\cite{Constan2013})\\\hline
         AB$_1$   & 3.31      & 0.21        & -0.21       & -1 (2\cite{Gilbert2019}, 9\cite{Gilbert2019})\\\hline
         AB$_2$   & 3.46      & 0.21        & -0.21       & 20 (18\cite{Gilbert2019}, 33\cite{Constan2013})\\\hline
    \end{tabular}
\end{table}

The main properties of the bilayer are presented in Table~\ref{tab:bilayer}. Energies align closely with those reported in other {\color{black}theoretical} studies.\cite{Gilbert2019, Ribeiro2011} In particular, the ground state corresponds to the AB structure instead of the AA' stacking, which is usually found in real samples. The results are not as similar to the calculations performed with wave function methods\cite{Constan2013}. Despite the high accuracy of these methods in molecules, it is not clear to the authors whether this applies to vdW crystals. The only notable difference from the DFT-based literature is a lower energy of AB$_1$ relative to  AA' stacking. Our methodology to obtain the total energy (HSE06) is expected to be more accurate than a PBE or vdW-corrected PBE scheme. However, this energy difference is less than 1 meV per atom and is not expected to be relevant for SPEs.

The interlayer distance aligns with previous calculations without a van der Waals correction \cite{Ribeiro2011,Constan2013}. Hirshfeld charges, indicative of the net charge at each atomic site, are independent of the stacking order. Therefore, the electrostatic potential exerted by one layer on the other depends solely on the \textit{local} environment and the distance between layers. Throughout the remainder, unless explicitly stated otherwise, we will use an interlayer distance of 3.34~\AA{} (AA' stacking, the most abundant one), isolating the effects on SPEs.

\subsection{Single photon emitters in the hBN bilayer}
\label{sec:spes}
{\color{black} 

Several luminescent defects have been found in hBN\cite{tran2016}. One particular type of SPEs are very bright and have a large dispersion in their ZPL that ranges from the near infrared to the near ultraviolet region.\cite{tran2016} Independent of the actual ZPL energy, these defects share a similar phonon sideband, with prominent phonon replicas at $\sim160-180$ meV.\cite{Wigger_2019} Photoluminescence spectra show a well-defined ZPL in addition to one or two high energy replicas.\cite{Martinez2016} These SPEs have a Huang-Rhys factor, $S_{HR}\approx 1$,\cite{Exarhos2017,winter2021} implying that about 1/3 of the measured photons belong to the central emission peak. In comparison to bulk defects, the NV$^-$ defect has $S_{HR}\approx 3.49$,\cite{Kurtsiefer2000} while the SiV$^0$ color center has a tiny coupling with the phonons, $S_{HR}=0.24$.\cite{Haussler_2017} Nearly 3\% and 80\% of the emitted photons correspond to the main line, respectively. Experiments associated these bright SPEs in hBN with C defects.\cite{mendelson2021} DFT calculations identified them as substitutional C defects, clusters of C defects, and more complex C-based defects, explaining the different emission energies and their characteristic phonon sideband.\cite{jara2021,pinilla2022} 
C-based defects form donor-acceptor-like systems, among the simplest and probably most abundant, we can mention dimer defects C$_N$C$_B$.\cite{Macia2022,Mack2019} They can be nearest neighbors, denoted C$_2$, or separated by some distance (C$_2$-$d$, see Ref.~\cite{Auburger21}), see Fig.~\ref{fig:stackHBN}-b. Also, we are interested in defects with an odd number of C atoms such as a single C$_N$ and trimers such as C$_N$C$_B$C$_N$ or simply C$_2$C$_N$.\cite{Li22} By considering this type of defect, the phonon sideband can be easily understood: excitation involves charge transfer, in a 2D lattice the attraction/repulsion translates into bond stretching, only activated in energetic optical phonon modes. 

\begin{figure}[h!]
    \centering
    \includegraphics[width=0.85\columnwidth]{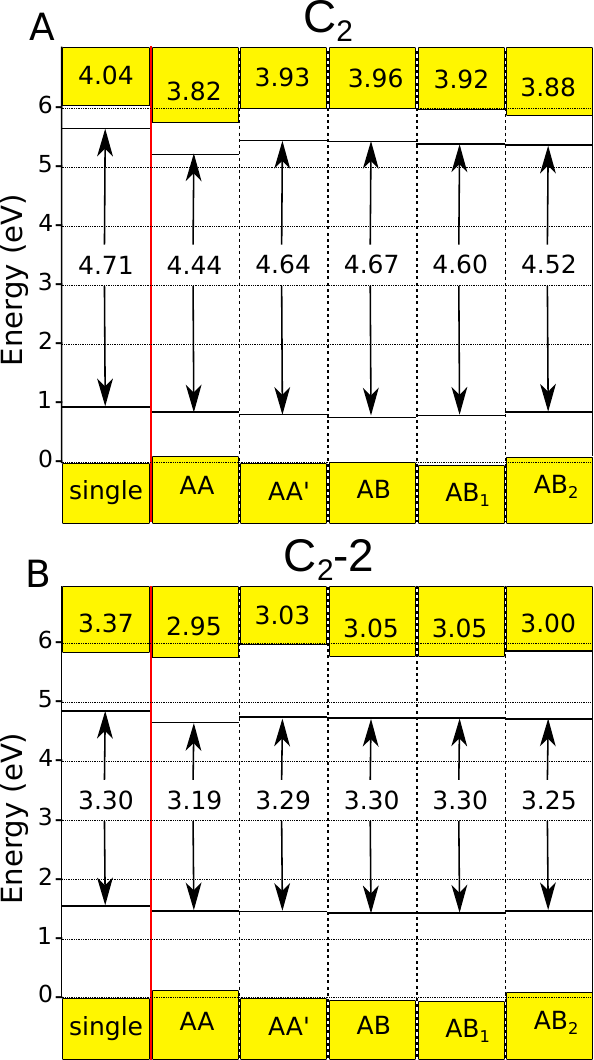}
    \caption{Energy levels for C$_2$ (panel A) and C$_2-2$ defects (panel B). Valence and conduction bands are denoted by yellow regions, with the stacking indicated in the lower yellow region. Defect levels are represented by solid black lines. The number between the arrows indicates the energy difference between the levels involved in the optical transition (the arrows point to these levels). Note that the single layer has a different reference energy than the bilayers (likely are shifted in energy). The energy scale was set to zero for the highest occupied bulk level of the single layer and AA' stacking. { \color{black}The changes in band edges of different bilayers are calculated with respect to the vacuum levels. The energy levels were calculated with DFT.} }
    \label{fig:C2}
\end{figure}

There exists another type of SPEs hosted in hBN, based on vacancies. They have a much lower brightness than C-based defects and are often measured as an ensemble.\cite{Baber2022} Among them, defects V$_B^-$  and C$_B$V$_N$ highlights because they are spin-active, with a triplet ground state,\cite{Ivady2020,CHENG2017247} just like the NV$^-$ center. In general, for vacancy-related defects in hBN, the ZPL line is a tiny fraction of the phonon sideband.\cite{jara2021} From a theoretical perspective, an accurate description of these defects often requires calculation methods more sophisticated than DFT.\cite{Reimers2018} Nevertheless, our calculation methods provide a qualitatively correct description of the defect C$_B$V$_N$

In this section, the results are exclusively from DFT calculations; as specified with the methods, these results will be explained in light of the qualitative insight of Eq.~\ref{eq:gap1}.}

The ground-state energy levels of defects C$_2$ and C$_2$-2, for each stacking, are illustrated in Fig.~\ref{fig:C2}, with other dimer defects showing similar behavior. {\color{black} The fundamental band gap of hBN is close to 6 eV, regardless of stacking or defects. Within it, there are two spin-degenerate defect levels, the lower one fully occupied and mostly localized in the C$_N$ atom. The defect level closer to the conduction band is unoccupied and is mostly localized in the C$_B$ atom, as in the scheme of Fig.~\ref{fig:potential}f. This electronic structure could be understood considering the different electronegativity (or on-site energy) of both C$_N$ and C$_B$ and the interaction between them. In C$_2$, as the defect atoms are adjacent, the interaction is quite strong, but it decreases quickly with separation in a large band-gap insulator. However, electronegativity does not depend on distance, keeping a finite defect energy gap even at the dissociation limit\cite{jara2021,Auburger21}.}

{\color{black}The energy levels of the C$_2$ defect show a marked difference between the hBN monolayer and the bilayer, for any stacking. This is an effect of enhanced screening in the bilayer.} Previous results, with more sophisticated methods, have shown a similar behavior.\cite{amblard2022universal} However, this enhancement of screening in multilayers is not significantly larger than in the bilayer. For non-adjacent defects, as exemplified by C$_2$-2 in Fig.~\ref{fig:C2}-B, the extra screening is less relevant, as expected due to the larger distance {\color{black} and weaker fields involved}.

Different staking orders in the hBN bilayer result in variations in the energy gap {\color{black} between defect levels. For defect C$_2$, this energy gap goes from 4.44 eV (AA stacking) to 4.67 eV (AB stacking); see Fig.~\ref{fig:C2}. That is, a modulation of up to $\sim 200$ meV in the defect band gap induced by changing the stacking of the bilayer. Instead, for the case of the defect C$_2$-2, the energy gap between the defect levels goes from 3.19 eV (AA stacking) to 3.30 eV in the AB stacking. This time, the energy gap was modulated $\sim 100$ meV by changing the stacking. For both defects, the stacking AA' results in a modulation of the defect gap very close to the stacking AB. Other defects studied follow a similar trend.

\begin{figure}[h]
    \centering
    \includegraphics[width=\columnwidth]{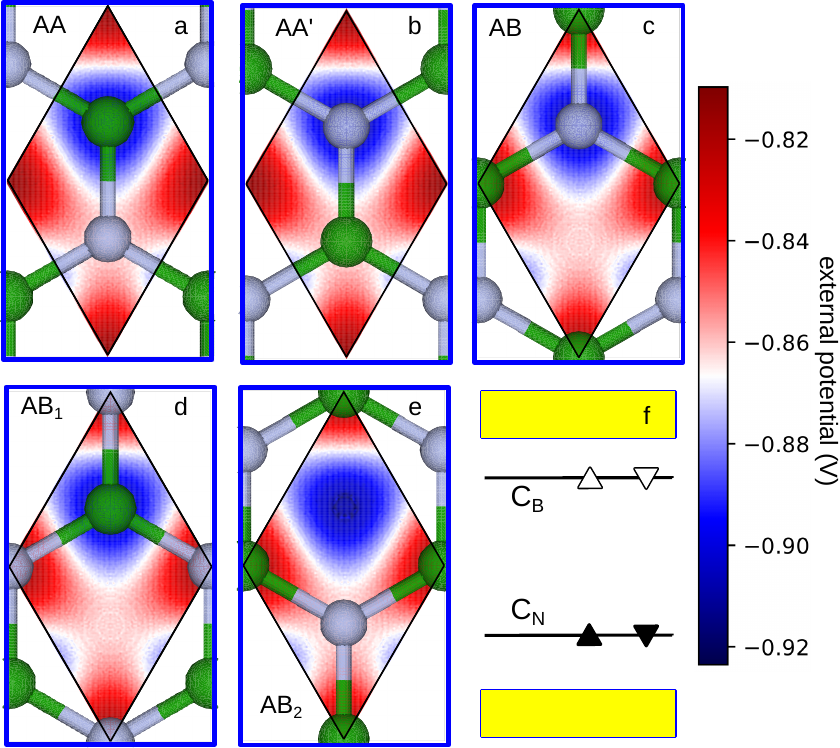}
    \caption{(a-e) Different stacking sequences: Only the atoms of the top layer are shown. The electrostatic potential applied by the bottom layer is displayed within the unit cell of the top layer, it is at the equilibrium distance of the AA' stacking. The colormap indicates the potential scale. {\color{black} C$_N$ and C$_B$ defects should replace the gray and green atoms, respectively.} (f) Energy levels of C$_N$-C$_B$ defect: In this scheme, the yellow regions represent bulk bands, and the black lines denote the $C_N$ and $C_B$ defect states. Filled (empty) triangles represent occupied (empty) states with different spins.}
    \label{fig:potential}
\end{figure}

The previous results can be qualitatively understood from the simple model of Eq.~\ref{eq:gap2}. Here $\Omega_H$ and $\Omega_L$ are the defects C$_N$ and C$_B$, respectively. As a first approximation, we can think of them as the positions of the N and B atoms in the hBN lattice, avoiding averaging the potential $\Phi(r)$. Then $\Delta\varepsilon_{HL-gap}$ is just the difference in the electrostatic potential between both lattice sites, $\Delta\varepsilon_{HL-gap}=\delta\Phi(r_{N})-\delta\Phi(r_{B})$. Fig.~\ref{fig:potential} shows the \textit{external} potential induced by one layer on the other for the different stackings studied. For the stacking AA, $\Phi(r_N)-\Phi(r_B)\approx -70$ meV, but for the stacking AA' the lattice sites are interchanged, giving $\Phi(r_N)-\Phi(r_B)\approx 70$ meV. Therefore, the defect gap in the AA' stacking should be $\approx 140$ meV larger than in the AA stacking, in good agreement with the DFT results of Fig.~\ref{fig:C2}a ($4.64-4.44$ eV $=0.2$ eV). Once again, we want to note that Eq.~\ref{eq:gap2} does not predict the gap between defect levels on a bilayer, but the relative changes due to changing the electrostatic potential (\textit{i.e.} the stacking).

For the defect C$_2$-2, and for any defect of type C$_2$-d, such a naive application of Eq.~\ref{eq:gap2} gives the same result as for C$_2$, $\Delta\varepsilon_{HL-gap}\approx 140$ meV when comparing the stacking AA and AA', this time the DFT value is underestimated. In this case, the inaccuracy comes from the fact that defects are taken as point charges. When the defects atoms are separated, part of the charge density is spread along the line that joins them\cite{pinilla2022}. This effect reduces $\Delta\varepsilon_{HL-gap}$, which is consistent with the DFT results.}

The changes of the defect gap also apply for the ZPL energies; see Table~\ref{tab:zpl}. {\color{black} The ZPL energies were calculated with DFT, simulating the excited level by setting the occupations manually.} The correspondence between the main emission energy and the energy gap of the defect levels is not complete. Although changes in the ZPL due to different stacking orders generally correlate with alterations in the defect gap, the degree of variation can be more or less pronounced for different defects.  This is because, in the excited state, the occupations may not follow the simple donor-acceptor picture employed in our model. 

A particularly interesting case is defect C$_N$. It has a ground state with its HOMO localized on the C atom. However, its LUMO (occupied upon excitation) is mainly in the B atoms close to $C_N$, with a node on the C atom. In other words, Eq.~\ref{eq:gap2} provides a great approximation, since $\Omega_H$ and $\Omega_L$ have minimal overlap, even in the excited state. This induces a larger change in ZPL when changing stacking order, as shown in Table~\ref{tab:zpl}.

\begin{table*}[h]
    \centering
    \caption{Zero phonon line of some luminescent defects in hBN, in eV. The system C$_2$-t refers to an hBN trilayer, the C$_2$ defect is in the middle layer.}
    \label{tab:zpl}

    \begin{tabular}{|cccccccc|}
        \hline
         Stacking  & C$_2$& C$_2$-2 & C$_2$-$\sqrt{7}$ & C$_N$ & C$_2$C$_N$ & C$_B$V$_N$ & C$_2$-t\\\hline
         AA        & 3.82  & 2.95  & 2.62  & 2.37  & 1.75 & 1.43 & 3.44 \\\hline
         AA'       & 3.93  & 3.03  & 2.72  & 2.52  & 1.76 & 1.82 & 3.66 \\\hline
         AB        & 3.96  & 3.05  & 2.71  & 2.59  & 1.76 & 1.70 & 3.71 \\\hline
         AB$_1$    & 3.92  & 3.05  & 2.71  & 2.55  & 1.76 & 1.67 & 3.64 \\\hline
         AB$_2$    & 3.88  & 3.00  & 2.67  & 2.37  & 1.75 & 1.63 & 3.54 \\\hline
         monolayer & 4.04  & 3.37  & 2.70  & 2.41  & 1.76 & 1.55 & 4.04 \\\hline
    \end{tabular}
\end{table*}

{\color{black}In stark contrast to other cases, the defect C$_2$C$_N$ presents the smallest differences in ZPL energy due to stacking order, up to 10 meV. In this case, the approximation of $\Omega_{H,L}$ as the sites of C$_N$, C$_B$ is incorrect; see Fig.~\ref{fig:trimer-wf}. For instance, the three C atoms have a relevant contribution to the LUMO, and it is not immediate to ignore the contribution of the N and B neighbors. Our aim is to provide a simple way to understand or even predict changes in the emission energy. Instead of calculating the actual interaction with the potential, Eq.~\ref{eq:gap1}. The projections of the wave function in atomic orbitals centered on each atom of the layer give $\int \left(|\phi_L|^2-|\phi_H|^2\right)dr=0.05$, confirming our previous observation.\cite{explanation}  

\begin{figure}[h]
    \centering
    \includegraphics[width=\columnwidth]{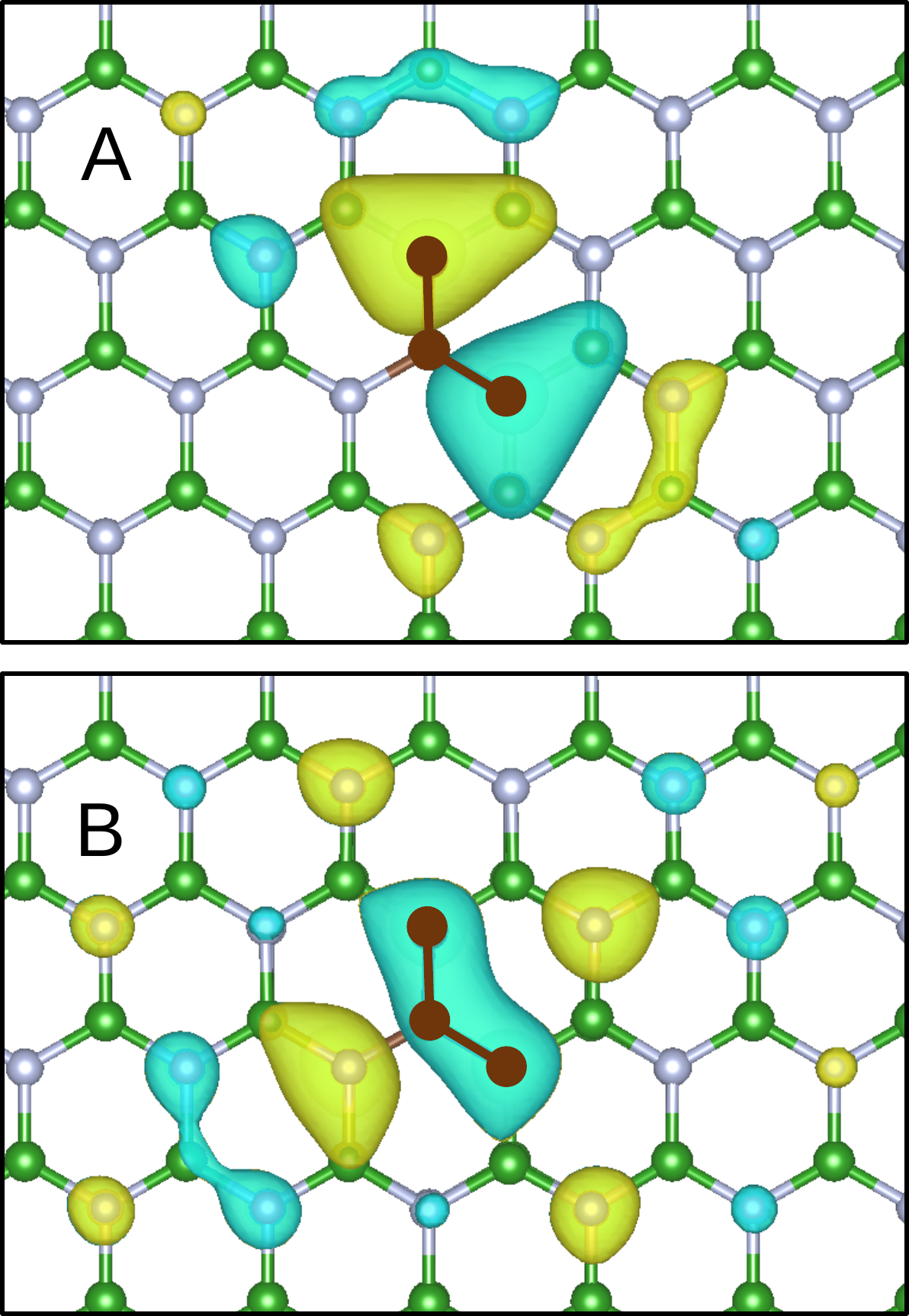}
    \caption{(a) HOMO and (b) LUMO of the defect C$_2$C$_N$. The different colors is the phase (positive or negative) of the wave function. A isovalue of $1\times10^{-7}$ was used. The C atoms are highlighted by brown circles. }
    \label{fig:trimer-wf}
\end{figure}

The only vacancy included in Table \ref{tab:zpl} is C$_B$V$_N$, the energy difference between the different stacking orders is up to 390 meV. This is more than three times what we could expect from Eq.\ref{eq:gap1}, even in the most favorable distribution of densities $|\phi_{L,H}|^2$. On the contrary, the wave functions of the relevant levels of C$_B$V$_N$ in the monolayer are distributed similarly between the B and N atoms, the worst scenario for changing the energy levels in Eq.~\ref{eq:gap1}. After inspecting the wave functions (or their projections), they are substantially affected by the stacking, changing them to have a greater participation of the B or N atoms according to the external potential. Although this invalidates Eq.~\ref{eq:gap1}, the effect is enhanced due to changes in HOMO and LUMO. 

The last case studied is a C$_2$ defect in a trilayer; the defect is embedded in the inner layer. In general, the ZPL decreases, again due to an increase in the screening (it should be close to its bulk value). The effect of the shift in the ZPL as a result of changes in the stacking is practically double with respect to the bilayer. This is expected because the external potential at the central layer is doubled. A further increase in the number of layers should have a marginal effect, since the potential from more distant layers should be very small. 
}

The changes in the ZPL induced by changes in stacking of a hBN bilayer are in the range of $\sim 10-390$ meV. For comparison, a `giant Stark effect' measured in SPEs hosted in hBN is on the order of 30 meV \cite{Noh2018}. The effect of the electrostatic potential due to different stacking in the hBN bilayer can be tested by sliding one layer with respect to the other. Nevertheless, it could be much larger in van der Waals heterostructures formed by hBN plus another 2D material with an in-plane dipolar texture. In particular, a 2D ferroelectric like In$_2$Se$_3$ could induce a substantial change in the ZPL controlled solely by a gate potential\cite{ding2017}. Such a study is beyond the scope of this contribution, as other effects beyond the screening and electrostatic interaction take place. 

{\color{black} Before closing this article, we need to comment on the effects of the bilayer in the phonon sideband of the defects studied. We did not find any relevant effect after moving from the monolayer to the bilayer or after changing the stacking order. That was expected, since the Franck-Condon relaxation resonates with optical phonons representing bond stretching at the defects and nearest neighbors. A change in screening or minor changes in electrostatic potential is far from sufficient to modify the strength of the chemical bonding in hBN. However, other defects, such as those based on vacancies, resonate with several low-energy breathing modes. In that case, it would not be clear that such a negligible effect would exist.}

\section{\label{sec:conclusions}Conclusions}
{\color{black}
This work presents a novel approach for manipulating single-photon emitters by harnessing electrostatic interaction within van der Waals heterostructures. Unlike conventional methods -strain, Stark effect- limited to $\sim 30$ meV tunability, our study using C-based defects in hexagonal boron nitride bilayers demonstrates shifts close to 400 meV for some defects, paving the way for highly tunable quantum light sources. This innovative strategy offers several advantages, including larger tunability, potential for on-chip integration, and greater applicability to various 2D emitters and heterostructures. 

Moreover, our findings are valid for most emitting defects as long as the levels involved in the transition are within the heterostructure band gap. This opens exciting avenues for exploring electrostatic control in diverse van der Waals heterostructures, potentially impacting quantum technologies like computing, communication, and sensing. 
}

\begin{acknowledgement}
We acknowledge financial support from FONDECYT through grants 1220366, 1231487, 1220715 and by Center for the Development of Nanosciences and Nanotechnology, CEDENNA AFB 220001. FM is supported by  Conicyt PIA/Anillo ACT192023. W.M. acknowledges the support of ANID Chile through the Doctoral National Scholarship N$^{\circ}\ 21211501$. J.C. acknowledges the support of ANID Chile through the Doctoral National Scholarship N$^{\circ}\ 21231429$ Powered@NLHPC: This research was partially supported by the supercomputing infrastructure of the NLHPC (ECM-02). 
\end{acknowledgement}



\bibliography{bib}

\end{document}